\documentclass{article}
\usepackage{spconf}
\usepackage{subcaption}
\usepackage{amsmath,amssymb}
\usepackage{hyperref}
\usepackage{url}
\usepackage{booktabs}       
\usepackage{amsfonts}       
\usepackage{nicefrac}       
\usepackage{microtype}      
\usepackage{xcolor}         
\usepackage[pdftex]{graphicx}
\usepackage[numbers, sort]{natbib}
\usepackage{tikz}
\usepackage{pgfplots, pgfplotstable}
\usepackage{enumitem}
\pgfplotsset{compat=1.18}
\usetikzlibrary{calc, shapes}
\usepackage{arydshln}
\usepackage{multirow}
\usepackage{xspace}
\usepackage{setspace} 
\usepackage{natbib}
\makeatletter
\DeclareRobustCommand\onedot{\futurelet\@let@token\@onedot}
\def\@onedot{\ifx\@let@token.\else.\null\fi\xspace}

\makeatother

\title{EnCLAP: Combining Neural Audio Codec and Audio-Text Joint Embedding for Automated Audio Captioning}
\name{Jaeyeon Kim$^{1,2}$ \qquad Jaeyoon Jung$^{1,3}$ \qquad Jinjoo Lee$^{1}$ \qquad Sang Hoon Woo$^{4}$
\address{
  $^{1}$MAUM AI Inc., Republic of Korea, \\ $^{2}$Seoul National University, Republic of Korea, \\ $^{3}$Soongsil University, Republic of Korea, \\ $^{4}$Independent Researcher}
\thanks{This work was supported by Institute of Information \& communications Technology Planning \& Evaluation (IITP) grant funded by the Korea government(MSIT) (No. 2021-0-00062-004, Development of Joint Work Automation Management Software Technology Based on Task Awareness)}
}

\begin{document}

\maketitle
\begin{abstract}
We propose EnCLAP, a novel framework for automated audio captioning. EnCLAP employs two acoustic representation models, EnCodec and CLAP, along with a pretrained language model, BART. We also introduce a new training objective called masked codec modeling that improves acoustic awareness of the pretrained language model. Experimental results on AudioCaps and Clotho demonstrate that our model surpasses the performance of baseline models. Source code will be available at \url{https://github.com/jaeyeonkim99/EnCLAP}. \footnote{An online demo is available at \url{https://huggingface.co/spaces/enclap-team/enclap}}
\end{abstract}

\begin{keywords}
automated audio captioning, neural audio codec, audio-text joint embedding
\end{keywords}

\section{Introduction}
Automated audio captioning (AAC) refers to the process of generating textual descriptions from audio signals that contain a wide range of sound events \cite{aac}. 
AAC can be classified as a cross-modal translation task from audio to natural language; thus, common approaches employ the encoder-decoder framework. The choice of architecture for the components varies between studies, encompassing CNN, RNN, and Transformer. 

Despite the recent advancements in neural networks, there still exists a substantial discrepancy between model performance and human performance \cite{audiocaps}. One significant factor of the discrepancy is the inherent complexity of the task. 
Moreover, the scarcity of data for AAC compared to its computer vision counterpart, exacerbates this issue. Specifically, AudioCaps \cite{audiocaps} and Clotho \cite{clotho}, two of the most widely used audio caption datasets, contain approximately 50K and 20K captions for training, respectively. In contrast, COCO captions \cite{coco_captions}, a widely used image caption dataset, has more than 414K captions for training.

In order to address this challenge, prior works have extensively used the transfer learning framework. Some works employed audio tagging or sound event detection as the pretraining task for the audio encoder \cite{mei, maac, localaft, graphac, convnext}. Others utilized pretrained language models such as GPT-2 \cite{gpt2, koizumi, prefix_tuning, pengi} and BART \cite{bart, gontier} as the text decoder to enhance the semantic quality of the captions. Some incorporated auxiliary losses, including keyword prediction loss \cite{koizumi_keyword} or sentence embedding loss \cite{sentence_embedding}, to improve the training procedure.

In this work, we present EnCLAP, a novel automated audio captioning framework. EnCLAP utilizes two pretrained acoustic feature encoders and a pretrained language model. Specifically, EnCLAP employs the CLAP \cite{clap_laion} audio encoder to compute sequence-level acoustic representations. For time step-level representations, we hypothesize that pretrained language models can leverage discrete inputs better compared to continuous inputs. Therefore, we use the EnCodec \cite{encodec} encoder, which yields discrete neural codec sequences, as the time step-level feature encoder. For the caption decoder, we finetune a pretrained BART \cite{bart} with an auxiliary training task to improve its acoustic awareness.
The contributions of this work can be summarized as follows:

\vspace{-1mm}
\begin{itemize}
    \setlength\itemsep{-1mm}
    \item EnCLAP achieves state-of-the-art performance on the AudioCaps dataset.
    \item We propose masked codec modeling, an auxiliary task designed to enhance the acoustic awareness of the BART encoder.
    \item Our ablation study demonstrates that the neural codec, when used in conjunction with sequence-level audio representation, is the more optimized input to pretrained language models compared to continuous inputs.
\end{itemize}
\vspace{-\baselineskip}
\begin{figure*}[t!]
    \centering
    \includegraphics[width=0.95\textwidth]{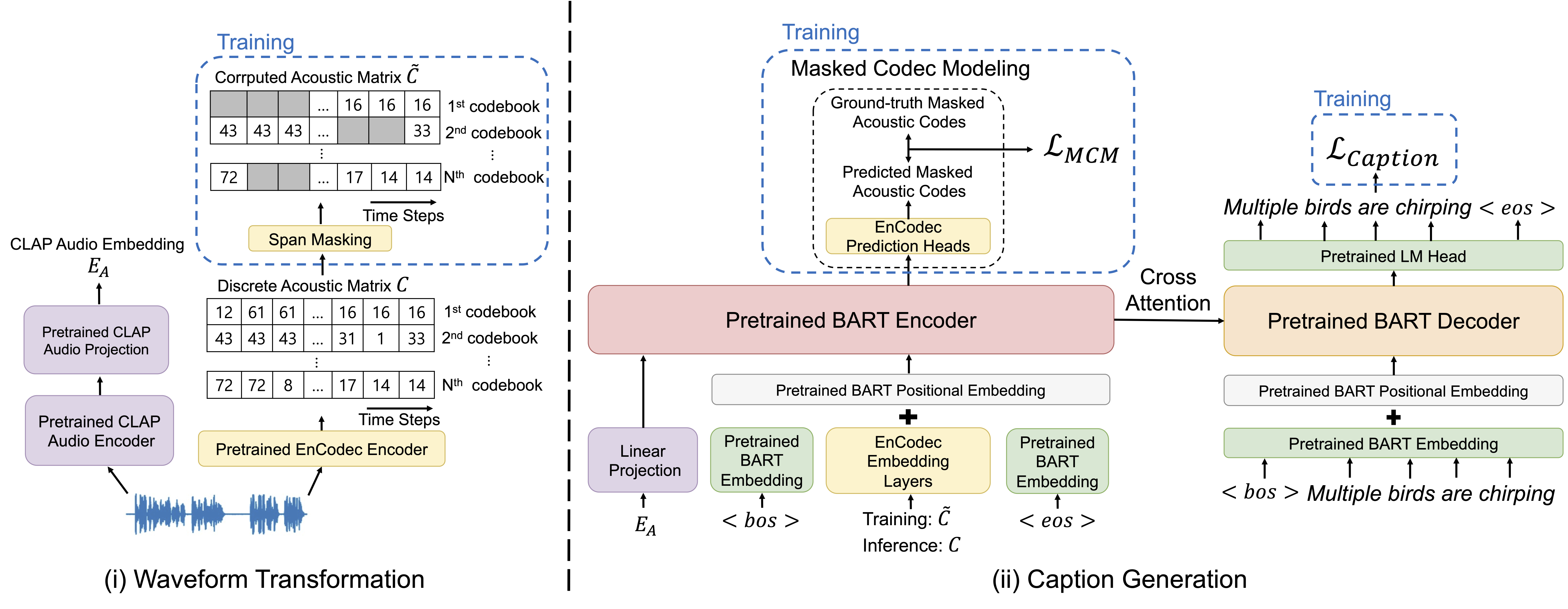}
    \vspace{-\baselineskip}
    \caption{The overall procedure of EnCLAP.} 
    \label{fig:enclap}
    \vspace{-\baselineskip}
\end{figure*}

\section{Proposed Method}
\subsection{EnCLAP}
Our proposed method, EnCLAP generates a caption from a given acoustic waveform in two stages. Initially, we process the waveform using two parallel encoders: the EnCodec \cite{encodec} encoder and the CLAP \cite{clap_laion} audio encoder. This process yields a discrete acoustic code matrix and a sequence-level audio embedding. Subsequently, we concatenate the two outputs and feed them to a pretrained BART model, which then generates the corresponding caption. An overview of the entire procedure is illustrated in Figure \ref{fig:enclap}.

\noindent\textbf{EnCodec. }
EnCodec \cite{encodec} is a neural codec model based on a convolutional auto-encoder. 
The encoder transforms an audio signal into parallel discrete acoustic code sequences using Residual Vector Quantization (RVQ) 
and the decoder reconstructs the waveform from the code sequences. Specifically, the EnCodec encoder maps a waveform to a discrete acoustic code matrix $C \in V^{N \times L}$, where $N$ is the number of codebooks used for RVQ, $L$ is the encoded audio length, and $V$ is the vocabulary of the codebooks. The row vector $c_{n,:}$ represents the code sequence from the $n$-th codebook. Conversely, the column vector $c_{:,l}$ comprises a code from each of the $N$ codebooks at time step $l$. RVQ utilizes a series of quantizers, where each quantizer encodes the quantization error from the previous quantizer.
We use $C$ as the input acoustic features.

\noindent\textbf{CLAP. }
Contrastive Language-Audio Pretraining (CLAP) \cite{clap, clap_laion} framework connects an audio and its text description into a joint multimodal space through dual encoders and contrastive learning. For an audio sample $A$ and the corresponding text description $T$, we compute their respective embeddings $E_A$ and $E_T$ using the following procedure: (i) $A$ and $T$ are independently processed through their specific encoders (ii) the encoder outputs are then pooled and mapped to a shared $D$-dimensional space using their respective projection layers.
The encoders and the projection layers are jointly trained using a contrastive objective, designed to maximize the similarity between embeddings of positive audio-text pairs while minimizing the similarity between embeddings of negative pairs. In this work, we only use the audio encoder and its projection. 

The contrastive learning encourages the audio embedding $E_A$ to be similar to textual embedding $E_T$ that represents the semantic content of the textual description. 
We use $E_A$ as the sequence-level semantic representation of A.

\noindent\textbf{BART. } We utilize a pretrained BART \cite{bart} model as the caption decoder. 
Each code sequence $c_{n, :}$ from a discrete acoustic code matrix $C \in V^{N \times L}$ is processed through the corresponding embedding layer $W_{n}$. The sum of these embeddings forms a BART input $e_{encodec} \in \mathbb{R}^{L \times D_{b}}$ where $D_b$ is the dimension size of the BART model.
On the other hand, the CLAP audio embedding $E_A$ is projected using a linear layer to form a BART input, $e_{clap} \in \mathbb{R}^{D_{b}}$. 

Let $e_{bos}$ and $e_{eos}$ be the BART embeddings for the special tokens $<$bos$>$ and $<$eos$>$ that indicate the beginning and the end of the sentence, and $e_{pos}$ be the BART positional embeddings. Then, the input to the BART encoder, $I_{b}$, is constructed in the following manner:
\begin{gather}
    I_{concat} = Concat(e_{bos}, e_{encodec}, e_{eos}) \in \mathbb{R}^{(L+2) \times D_{b}} \\ 
    I_{pos} = I_{concat} + e_{pos} \in \mathbb{R}^{(L+2) \times D_{b}} \\
    I_{b} = Concat(e_{clap}, I_{pos}) \in \mathbb{R}^{(L+3) \times D_{b}}
\end{gather}  
BART autoregressively generates a caption, conditioned on $I_{b}$.

\vspace{-0.3\baselineskip}
\subsection{Training}
\vspace{-0.2\baselineskip}
We apply cross-entropy loss between the ground-truth caption and the generated caption. Specifically, for an audio $A$ and its caption $T$, the objective of the captioning task, which is to maximize the probability $p(T|A)$, can be optimized through the following objective function
\begin{gather}
    \mathcal{L}_{caption} = -\frac{1}{L_{T}}\sum_{t=1}^{L_{T}}logp(y_t|y_{1:t-1}, E_A, C)
\end{gather}
where $L_{T}$ is the length of $T$, and $y_t$ is a $t$-th token of $T$. The CLAP and the EnCodec models are frozen during training.

\vspace{-0.6\baselineskip}
\subsubsection{Masked Codec Modeling}
\vspace{-0.1\baselineskip}
We introduce masked codec modeling (MCM) as an auxiliary training task in order to guide the BART encoder to learn the contextualized relationships among acoustic codes. Given a discrete acoustic code matrix $C$, we apply span masking to each code sequence $c_{n,:}$ using a special mask token. We use the corrupted code matrix $\Tilde{C}$ in place of $C$ as the input to the BART model. Then, we add $N$ linear classifiers on top of the BART encoder, where $N$ is the number of codebooks in $C$. The $n$-th classifier predicts the ground-truth codes for the masked codes within $\Tilde{c}_{n,:}$. We compute the cross entropy loss between the ground-truth codes and the predicted codes.

The objective function of MCM is a sum of prediction losses across all code sequences. However, the loss from each code sequence is scaled differently, since the earlier codebooks carry more significance in RVQ. Specifically, the MCM objective is defined as
\vspace{-1mm}
\begin{gather}
    \mathcal{L}_{mcm} = -\sum_{i=1}^{N}(\frac{1}{2})^i \times logp(c_{i, :}|\Tilde{c}_{i, :})
\end{gather}
The modified training objective is:
\begin{gather}
\mathcal{L}_{total} = \mathcal{L}_{caption} + \lambda \times \mathcal{L}_{mcm} 
\end{gather}
\vspace{-1mm}
where $\lambda$ is a scaling hyperparameter between 0 and 1.
\section{Experiments}
\subsection{Setup}
\textbf{Datasets.} We train and evaluate EnCLAP on two most widely used AAC datasets, AudioCaps \cite{audiocaps} and Clotho \cite{clotho}.
AudioCaps is a subset of AudioSet \cite{audioset} that has been reannotated with caption labels. Each audio clip is annotated with a single caption for the training set and five captions for the validation and test sets. The dataset is divided into training, validation, and test sets containing 49838, 495, and 975 examples, respectively.
Clotho consists of audio clips sourced from Freesound, each labeled with 5 captions. In our experiment, we use version 2.1 of Clotho, which contains 3839 examples in the training set, 1045 in the validation set, and 1045 in the test set.

For AudioCaps, we train and evaluate using the original dataset split. For Clotho, we conduct the evaluation using two commonly used setups. In the first setup, we train a model exclusively on Clotho. In the second setup, we use a model pretrained on AudioCaps and finetune it on Clotho.

\noindent\textbf{Model Configuration. }
In our experiment, we use the pretrained CLAP model trained on LAION-Audio-630k \cite{clap_laion} and AudioSet \cite{audioset}. Additionally, we use the pretrained EnCodec model that translates a 24kHZ waveform to 16 discrete code sequences at the frequency of 75Hz, with each code assuming one of 1024 values. We present two versions of EnCLAP, EnCLAP-base and EnCLAP-large, based on the version of BART used. 

\noindent\textbf{Basline Models. } 
We compare EnCLAP to AAC models from prior works. For the AudioCaps dataset, we select state-of-the-art models for comparison. ACT \cite{mei} employs a vision transformer encoder along with a transformer decoder. LHDFF \cite{lhdff} leverages the frequently overlooked low-dimensional audio features by fusing them with the high-dimensional audio features. CNN14-GPT2 \cite{prefix_tuning} and Pengi \cite{pengi} encode the audio features and use them as prefixes to GPT2 \cite{gpt2}. The BART-tags \cite{gontier} model generates a caption conditioned on a sequence of predicted AudioSet \cite{audioset} tags, using a pretrained BART. CNeXt-Trans \cite{convnext} employs a ConvNeXt \cite{convnext_original_paper} encoder pretrained on audio classification along with a transformer decoder. AL-MixGen \cite{multitta} is ACT trained using audio-language mixup augmentation and evaluated with test-time augmentation. WavCaps \cite{wavcaps} is a HTSAT-BART \cite{hts-at} model trained on a large-scale weakly-labeled dataset that achieved state-of-the-art performance on AudioCaps.

Since the Clotho dataset is predominantly utilized for DCASE Challenge, we select a subset of models evaluated on Clotho as the baseline models to ensure a fair comparison. Specifically, we exclude any works that employ challenge-specific optimizations. These include the use of an ensemble of models or the application of reinforcement learning for metric optimizations. CLIP-AAC \cite{clip_aac} adds a text encoder to the conventional encoder-decoder architecture for contrastive learning. MAAC \cite{maac} employs an LSTM-based multimodal attention decoder to incorporate both the acoustic and the semantic information.  P-LocalAFT \cite{localaft} employs CNN-10 from PANN \cite{pann} as the encoder and an local information assisted attention-free transformer as the decoder. GraphAC \cite{graphac} introduces a graph attention module after the PANN encoder.

\noindent\textbf{Evaluation Metrics.} We adopt widely used AAC metrics METEOR \cite{meteor}, CIDEr \cite{cider}, SPICE \cite{spice}, and SPIDEr \cite{spider} as the evaluation metrics. All metrics are calculated using the aac-metrics library.\footnote{\url{https://pypi.org/project/aac-metrics/}} METEOR is a machine translation evaluation metric, based on unigram precision and recall. CIDEr and SPICE assess the syntactic and semantic quality of the generated captions, respectively. SPIDEr is a linear combination of SPICE and CIDEr, designed to capture both syntactic and semantic aspects of the caption.

\noindent\textbf{Training Details.} We train EnCLAP models on Nvidia A100 GPUs for $15$ epochs. We use AdamW optimizer with $\beta_1 = 0.9$, $\beta_2 = 0.999$, and weight decay coefficient of $0.01$.
We apply a label smoothing factor of 0.2 only for caption generation.
We use inverse square root learning rate scheduler 
with linear warm-up. For AudioCaps, we apply warm-up for 2000 steps until it reaches the peak learning rate of 6.5e-5 and 3e-5 for the base model and the large model, respectively. Similarly. for Clotho, we warm up for 1000 steps to the peak learning rate, which is 4e-5 for the base model and  2.5e-5 for the large model. For finetuning an AudioCaps model on Clotho, we use the inverse square root scheduler without warm-up, with initial learning rates of 2e-5 for the base model and 1.25e-5 for the large model.
For MCM, we mask 15\% of each codebook sequence. The length of each masking span is set to 10. 
We use 0.7 for MCM loss scaling factor $\lambda$.

\vspace{-0.5\baselineskip}
\subsection{Results}
\textbf{Results on AudioCaps.} We provide the 
results on the AudioCaps dataset in Table~\ref{table:audiocaps}. EnCLAP-base models outperform most baseline models and matches the performance of WavCaps \cite{wavcaps}. 
Notably, EnCLAP-large establishes the new start-of-the-art result. 
Despite being trained solely on the original AudioCaps dataset, EnCLAP-large surpasses even the WavCaps model, which was pretrained on a large-scale dataset.

\begin{table}[t]
\caption{Results on AudioCaps.}
\vspace{-0.5\baselineskip}
\label{table:audiocaps}
\centering
\resizebox{\linewidth}{!}{
\begin{tabular}{ccccc}
\hline
Model & METEOR ($\uparrow$) & CIDEr ($\uparrow$) & SPICE ($\uparrow$) & SPIDEr ($\uparrow$)\\
\hline
ACT \cite{mei} & 0.222 & 0.679 & 0.160 & 0.420 \\
LHDFF \cite{lhdff} & 0.232 & 0.680 & 0.171 & 0.426 \\
CNN14-GPT2 \cite{prefix_tuning} & 0.240 &  0.733 & 0.177 & 0.455 \\
BART-tags \cite{gontier} & 0.241 & 0.753 & 0.176 & 0.465 \\
Pengi \cite{pengi} & 0.232 & 0.752 & 0.182 & 0.467 \\ 

CNeXt-Trans \cite{convnext} & - & 0.760 & 0.182 & 0.471 \\
AL-MixGen \cite{multitta} & 0.242 & 0.769 & 0.181 & 0.475 \\
WavCaps \cite{wavcaps} &  0.250 & 0.787 & 0.182 & 0.485  \\
\cdashline{0-4}
EnCLAP-base & 0.247 & 0.780 & 0.186 & 0.483 \\
EnCLAP-large & \textbf{0.255} & \textbf{0.803} & \textbf{0.188} & \textbf{0.495} \\
\hline
\end{tabular}
\vspace{-0.8\baselineskip}
}
\end{table}

\begin{table}[t]
\caption{Results on Clotho.}
\vspace{-0.5\baselineskip}
\label{table:clotho}
\centering
\resizebox{\linewidth}{!}{
\begin{tabular}{ccccc}
\hline
Model & METEOR ($\uparrow$) & CIDEr ($\uparrow$) & SPICE ($\uparrow$) & SPIDEr ($\uparrow$)\\
\hline
\multicolumn{5}{c}{\textit{Trained on Clotho only}} \\
CLIP-AAC \cite{clip_aac} & 0.171 & 0.407 & 0.119 & 0.263 \\
LHDFF \cite{lhdff} & 0.175 & 0.408 & 0.122 & 0.265 \\
MAAC \cite{maac} & 0.174 & 0.419 & 0.119 & 0.269  \\
\cdashline{0-4}
EnCLAP-base & 0.180 & \textbf{0.461} & 0.128 & \textbf{0.294} \\
EnCLAP-large & \textbf{0.182} & 0.426 & \textbf{0.129} & 0.278 \\
\hline
\multicolumn{5}{c}{\textit{Trained on AudioCaps and Clotho}} \\
Pengi \cite{pengi} & 0.172 & 0.416 & 0.126 & 0.271\\
P-LocalAFT \cite{localaft} & 0.177 & 0.434 & 0.122 & 0.278 \\
GraphAC \cite{graphac} & 0.175 & 0.437 & 0.126 & 0.281 \\
\cdashline{0-4}
EnCLAP-base & 0.184 & 0.463 & 0.128 & 0.295 \\
EnCLAP-large & \textbf{0.186} & \textbf{0.464} & \textbf{0.133} & \textbf{0.299} \\
\hline
\end{tabular}
\vspace{-0.5\baselineskip}
}
\end{table}

\noindent\textbf{Result on Clotho.} Table~\ref{table:clotho} shows the evaluation results on the Clotho dataset. Under both setups, EnCLAP models exceed the baseline models. We observe that the performance of EnCLAP models trained on both AudioCaps and Clotho surpasses that of EnCLAP models trained solely on the Clotho dataset. We also note the inconsistent performance of EnCLAP-large; it underperforms EnCLAP-base when trained solely on Clotho, yet outperforms EnCLAP-base when trained on both AudioCaps and Clotho. Given the relatively small size of Clotho, we hypothesize that EnCLAP-large is more prone to overfitting in low-resource settings and may benefit from a more carefully designed regularization.

\begin{table}[t]
\caption{Ablation Results on Clotho.}
\vspace{-0.5\baselineskip}
\label{table:ablation}
\centering
\resizebox{\linewidth}{!}{
\begin{tabular}{lcccc}
\hline
Model & METEOR ($\uparrow$) & CIDEr ($\uparrow$) & SPICE ($\uparrow$) & SPIDEr ($\uparrow$) \\
\hline
\multicolumn{5}{c}{\textbf{Ablation on Key Components}} \\
EnCLAP & \textbf{0.180} & \textbf{0.461} & 0.128 & \textbf{0.294}\\
\quad - BART Pretrain & 0.171 & 0.401 & 0.122 & 0.262  \\
\quad - MCM & \textbf{0.180} & 0.440 & \textbf{0.130} & 0.285  \\
\quad\quad - EnCodec & 0.158 & 0.322 & 0.102 & 0.212 \\
\quad\quad - CLAP & 0.131 & 0.209 & 0.074  & 0.141 \\ 
\quad - CLAP  & 0.129 & 0.214 &  0.075 & 0.145 \\
\hline
\multicolumn{5}{c}{\textbf{Ablation on Acoustic Feature Representations}} \\
\multicolumn{5}{c}{\textit{Without CLAP}} \\
Log-Mel. & 0.120 & 0.183 & 0.068 & 0.126  \\
CNN-14 & \textbf{0.173} & \textbf{0.415} & \textbf{0.123} & \textbf{0.269} \\
EnCodec & 0.131 & 0.209 & 0.074  & 0.141 \\ 
\multicolumn{5}{c}{\textit{With CLAP}} \\
CLAP + Log-Mel. & 0.179 & 0.435 & 0.125 & 0.280 \\
CLAP + CNN-14 & 0.174 & 0.403 & 0.121 & 0.262 \\
CLAP + EnCodec & \textbf{0.180} & \textbf{0.440} & \textbf{0.130} & \textbf{0.285} \\
\hline
\end{tabular}
\vspace{-\baselineskip}
}
\end{table}
\subsection{Ablation}
We conduct ablation studies on the key components and the choice of acoustic feature representations. The experiment details follow that of EnCLAP-base on Clotho.

\noindent\textbf{Key Components}
We conduct an ablation experiment on the key components of EnCLAP, in particular the CLAP audio embedding, EnCodec, MCM training, and the pretrained BART. Table 3 demonstrates that the inclusion of each component contributes towards EnCLAP's performance. Notably, the CLAP audio embedding yields the most significant improvement.

\noindent\textbf{Acoustic Feature Representations. }
Additionally, we conduct an experiment to compare EnCodec to other widely-used acoustic feature representations in AAC. Specifically, our experiment includes EnCodec, the log-mel spectrogram and the output of the CNN-14 encoder from PANN \cite{pann}. For the log-mel spectrogram, we first perform short-time Fourier transform with Hamming window with a size of 40 milliseconds and a hop size of 20 milliseconds, and apply a mel filterbank of size 64. CNN-14 is the pretrained encoder from PANN that has been finetuned for audio retrieval.\footnote{\url{https://github.com/xieh97/dcase2023-audio-retrieval}} These representations are linearly projected before being fed to the BART encoder. We run the experiment under two setups: one in which the representations were used on their own, and the other where the representations were used in conjunction with the CLAP audio embeddings. 
Table~\ref{table:ablation} shows that EnCodec outperforms its alternatives in the setup with CLAP, while the CNN-14 baseline prevails without CLAP, likely because it captures global information unlike EnCodec or log-mel spectrograms. The lack of performance improvement of CNN-14 with CLAP further reinforces this hypothesis, suggesting that discrete neural codecs, augmented with sequence-level information, are the more optimal input format.

\section{Conclusion}
In this work, we presented a new framework for automated audio captioning called EnCLAP. EnCLAP comprises the CLAP audio encoder, EnCodec encoder, and the pretrained BART model. We also introduce masked codec modeling, an auxiliary task designed to improve the acoustic-awareness of a pretrained language model. Experiments show that our proposed method exhibits state-of-the-art performance in audio captioning. In future works, we plan to expand the application of EnCLAP to a wider range of tasks, including music captioning and audio generation.
\vfill\pagebreak
\bibliographystyle{IEEEbib}

\begin{spacing}{0.89}
\setlength{\bibsep}{3.5pt}
\bibliography{main}
\end{spacing}
\end{document}